\documentclass[conference,10pt,twocolumn]{IEEEtran}
\usepackage{graphicx}
\usepackage{multibib}
\usepackage{multirow}
\usepackage{epstopdf}
\usepackage{amsmath,amssymb,amsbsy}
\usepackage{subfigure,url,float,cite}
\usepackage[T1]{fontenc}
\usepackage[utf8]{inputenc}
\usepackage{graphicx}
\usepackage{multibib}
\usepackage{multirow}
\usepackage{epstopdf}
\usepackage{dsfont}
\usepackage[T1]{fontenc}
\usepackage[utf8]{inputenc}

\newtheorem{lemma}{Lemma}
\newtheorem{thm}{Theorem}

\usepackage{flushend}

\usepackage[ruled,lined,linesnumbered,noend]{algorithm2e}

\usepackage{comment}

\usepackage{color}

\usepackage{subfigure,float}

\DeclareMathOperator*{\argmax}{argmax}
\DeclareMathOperator*{\op}{\mathtt{op}}
\DeclareMathOperator*{\prd}{\mathtt{pred}}
\DeclareMathOperator*{\scc}{\mathtt{succ}}
\DeclareMathOperator*{\reals}{\mathbb{R}}

\newcommand{\jobset}{\ensuremath{\mathcal{G}}}
\newcommand{\nodeset}{\ensuremath{\mathcal{V}}}

\newcommand{\feasibledomain}{\ensuremath{\mathcal{D}}}
\newcommand{\id}{\ensuremath{\mathds{1}}}

\newcommand{\intechreport}[2]{#1} 

\newcommand{\prob}{\mathbf{P}}
\newcommand{\expect}{\mathbb{E}}
\IEEEoverridecommandlockouts
\begin{document}

\date{}

\title{\huge  Intermediate Data Caching Optimization for Multi-Stage and Parallel Big Data Frameworks \small{\thanks{This work was supported by CNS-1452751, CNS-1552525, CCF-1750539, and NeTS-1718355. }}}

%


\author{
	\IEEEauthorblockN{
		{Zhengyu Yang}\IEEEauthorrefmark{1},
		{Danlin Jia}\IEEEauthorrefmark{1},			
		{Stratis Ioannidis}\IEEEauthorrefmark{1},	
	    {Ningfang Mi}\IEEEauthorrefmark{1},
		and {Bo Sheng}\IEEEauthorrefmark{2}}	
	\IEEEauthorblockA{\IEEEauthorrefmark{1}Dept. of Electrical \& Computer
		Engineering, Northeastern University, 360 Huntington Ave., Boston, MA 02115
	}
	
	\IEEEauthorblockA{\IEEEauthorrefmark{2}Dept. of Computer Science,
		University of Massachusetts Boston, 100 Morrissey Boulevard, Boston, MA 02125\\
	     \{yang.zhe, jia.da\}@husky.neu.edu, \{ioannidis, ningfang\}@ece.neu.edu, shengbo@cs.umb.edu
	}		
}


\maketitle

\vspace{-2mm}

\begin{abstract}
In the era of big data and cloud computing, large amounts of data  are generated from  user applications and need to be processed in the datacenter.  
Data-parallel computing frameworks, such as Apache Spark, are widely used to perform such data processing at scale. 
Specifically, Spark  leverages distributed memory to cache the intermediate results, represented as Resilient Distributed Datasets (RDDs). This gives Spark an advantage over other parallel frameworks for implementations of iterative machine learning and data mining algorithms, by  avoiding repeated computation or hard disk accesses to retrieve RDDs.
%
By default, caching decisions are left at the programmer's discretion, and the LRU policy  is used for evicting RDDs when the cache is full. 
However, when the objective is to minimize total work, LRU is woefully inadequate, leading to arbitrarily suboptimal caching decisions. 
In this paper, we design an algorithm for multi-stage big data processing platforms to adaptively determine and cache the most valuable intermediate datasets that can be reused in the future. 
Our solution automates the decision of which RDDs to cache: this amounts to identifying nodes in a direct acyclic graph (DAG) representing computations whose outputs should persist in the memory. 
Our experiment results show that our proposed cache optimization solution can improve the performance of machine learning applications on Spark decreasing the total work to recompute RDDs by 12\%.
\end{abstract}

\vspace{-1mm}

\begin{keywords}
Cache Optimization, Multi-stage Framework, Intermediate Data Overlapping, Spark
\end{keywords}

\vspace{-3mm}
\section{Introduction}
\label{SEC:IN}
\vspace{-2mm}

With the rise of big data analytics and cloud computing, cluster-based large-scale data processing has become a common paradigm in many applications and services. 
Online companies of diverse sizes, ranging from technology giants to smaller startups, routinely store and process data generated by their users and applications on the cloud. 
Data-parallel computing frameworks, such as Apache Spark~\cite{zaharia2010spark,spark} and Hadoop~\cite{hadoop}, are employed to perform such data processing at scale. 
Jobs executed over such frameworks comprise hundreds or thousands of identical parallel subtasks, operating over massive datasets, and executed concurrently in a cluster environment.

\begin{figure}
\centering
\includegraphics[width=0.34\textwidth]{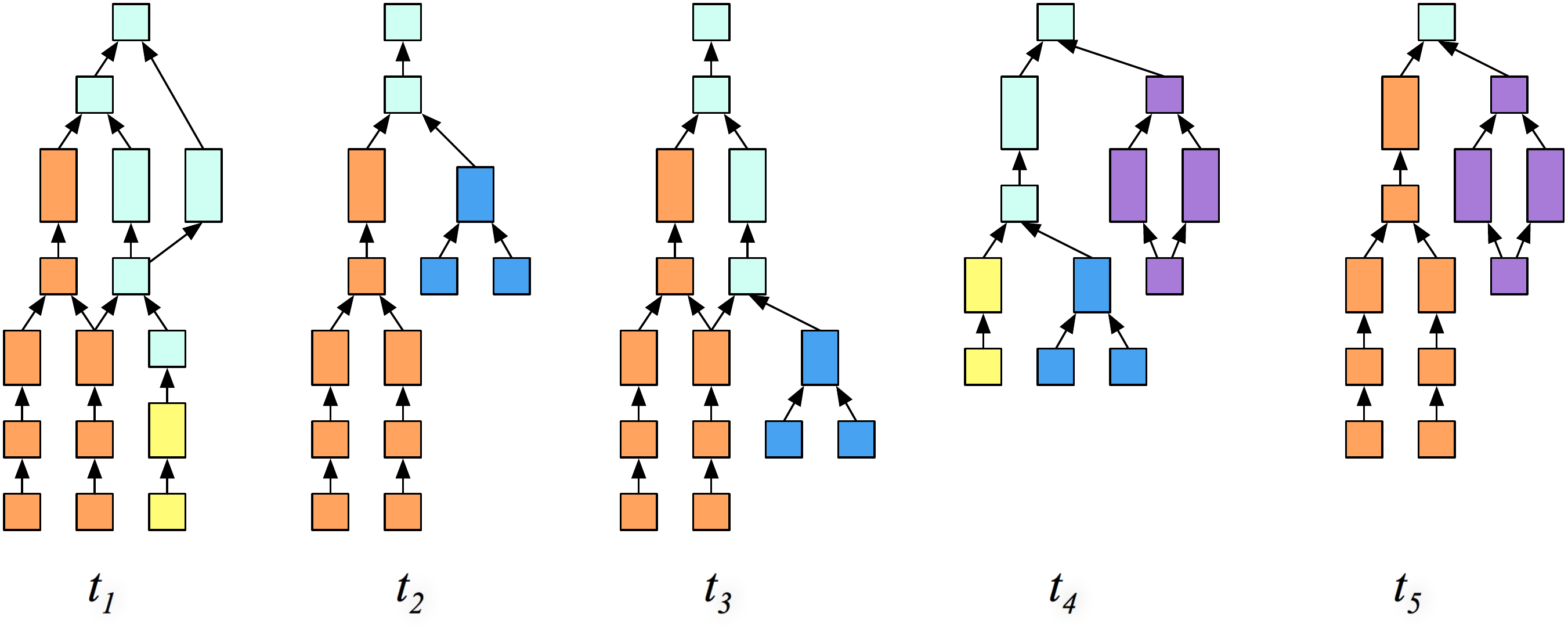}\vspace{-3mm}
\caption{\scriptsize {\bf{Job arrivals with computational overlaps.}} Jobs to be executed over the cluster arrive at different times $t_1,\ldots,t_5$. Each job is represented by a Directed Acyclic Graph (DAG), whose nodes correspond to operations, e.g., map, reduce, or join, while arrows represent order of precedence. 
Crucially, jobs have \emph{computational overlaps}: their DAGs comprise common sets of operations executed over the same data, indicated as subgraphs colored identically across different jobs. Caching such results can significantly reduce computation time.\vspace*{-1em}}
\label{FIG:JOBARRIVALS}
\end{figure}

\vspace{-1mm}
The time and resources necessary to process such massive jobs are immense. Nevertheless, jobs executed in such distributed environments often have significant computational overlaps: 
different jobs processing the same data may involve common intermediate computations, as illustrated in Fig.~\ref{FIG:JOBARRIVALS}.
Such computational overlaps arise naturally in practice. 
Indeed, computations performed by companies are often applied to the same data-pipeline: companies collect data generated by their applications and users, and store it in the cloud. 
Subsequent operations operate over the same pool of data, e.g., user data collected within the past few days or weeks. 
More importantly, a variety of prominent data mining and machine learning operations involve common preprocessing steps. This includes database projection and selection~\cite{maier1983theory}, preprocessing in supervised learning~\cite{trevor2001elements}, and dimensionality reduction~\cite{eldar2012compressed}, to name a few. 
Recent data traces from industry have reported $40\sim60\%$ recurring jobs in Microsoft production clusters~\cite{jyothi2016morpheus}, and up to $78\%$ jobs in Cloudera clusters involve data re-access~\cite{chen2012interactive}.

\vspace{-1mm}
Exploiting such computational overlaps has a tremendous potential to drastically reduce job computation costs and lead to significant performance improvements. In data-parallel computing frameworks like Spark, computational overlaps inside each job are exploited through caching and \emph{memoization}: the outcomes of computations are stored with the explicit purpose of significantly reducing the cost of subsequent jobs. 
On the other hand, introducing caching also gives rise to novel challenges in resource management;  
to that end, the purpose of this paper is to design, implement and evaluate caching algorithms over data-parallel cluster computing environments.

\vspace{-1mm}
Existing data-parallel computing frameworks, such as Spark, incorporate caching capabilities in their framework in a non-automated fashion. 
The decision of which computation results to cache rests on the developer that submits jobs: the developer explicitly states which results are to be cached, while cache eviction is implemented with the simple policy (e.g., LRU or FIFO); neither caching decisions nor evictions are part of an optimized design. Crucially, determining which outcomes to cache is a hard problem when dealing with jobs that consist of operations with complex dependencies. 
Indeed, under the Directed Acyclic Graph (DAG) structures illustrated in Fig.~\ref{FIG:JOBARRIVALS}, making caching decisions that minimize, e.g., total work is NP-hard~\cite{ioannidis2016adaptive,shanmugam2013femtocaching}. 

\vspace{-1mm}
In this paper, we develop an adaptive algorithm for caching in a massively distributed data-parallel cluster computing environment, handling complex and massive data flows. Specifically, a mathematical model is proposed for determining caching decisions that minimize total work, i.e., the total computation cost of a job. 
Under this mathematical model, 
we have developed new {\em adaptive} caching algorithms to  make online caching decisions with optimality guarantees, e.g., minimizing total execution time. 
%
Moreover, we  extensively validate the performance over several different databases, machine learning, and data mining patterns of traffic, both through simulations and through  an implementation over Spark, comparing and assessing their performance with respect to existing popular caching and scheduling policies.

The remainder of this paper is organized as follows. 
Sec.~\ref{SEC:BM} introduces background and motivation. 
Sec.~\ref{SEC:AD}  presents our model, problem formulation,  and   our proposed algorithms. 
Their performance is evaluated in Sec.~\ref{SEC:PE}. 
Sec.~\ref{SEC:RW} reviews related work, and we
conclude in Sec.~\ref{SEC:CO}.

\section{Background and Motivation}
\label{SEC:BM}
\vspace{-2mm}

\subsection{Resilient Distributed Datasets in Spark}
\vspace{-1mm}

Apache Spark has recently been gaining ground as an alternative for distributed data processing platforms. In contrast to Hadoop and MapReduce~\cite{dean2008mapreduce}, 
Spark is a memory-based general parallel computing framework. It provides {\em resilient distributed datasets} (RDDs) as a primary abstraction: RDDs are distributed datasets stored in RAM across multiple nodes in the cluster. 
In Spark, the decision of which RDDs to store in the RAM-based cache rests with the developer~\cite{zaharia2012resilient}: the developer explicitly requests for certain results to persist in RAM. Once the RAM cache is full, RDDs are evicted using the LRU  policy. Alternatively, developers are further given the option to store evicted RDDs on HDFS, at the additional cost of performing write operations on HDFS. RDDs cached in RAM are stored and retrieved faster; however, cache misses occur either because an RDD is not explicitly cached by the developer, or because it was cached and later evicted. In either case, Spark is resilient to misses at a significant computational overhead: if a requested RDD is neither in RAM nor stored in HDFS, Spark recomputes it from scratch. Overall, cache misses, therefore, incur additional latency, either by reading from HDFS or by fully recomputing the missing RDD. 

\vspace{-2mm}
An example of a job in a data-parallel computing framework like Spark is given in Fig.~\ref{FIG:DAG}. A job is represented as a DAG (sometimes referred to as the \emph{dependency graph}). Each node of the DAG corresponds to a parallel operation, such as reading a text file and distributing it across the cluster, or performing a map, reduce, or join operation. Edges in the DAG indicate the order of precedence: an operation cannot be executed before all operations pointing towards it are completed, because their outputs are used as inputs for this operation. As in existing frameworks like Spark or Hadoop, the inputs and outputs of operations may be distributed across multiple machines: e.g., the input and output of a map would be an RDD in Spark, or a file partitioned across multiple disks in HDFS in Hadoop.


\vspace{-3mm}
\subsection{Computational Overlaps}
\vspace{-2mm}

\begin{figure}[t]
\centering
\includegraphics[width=0.34\textwidth]{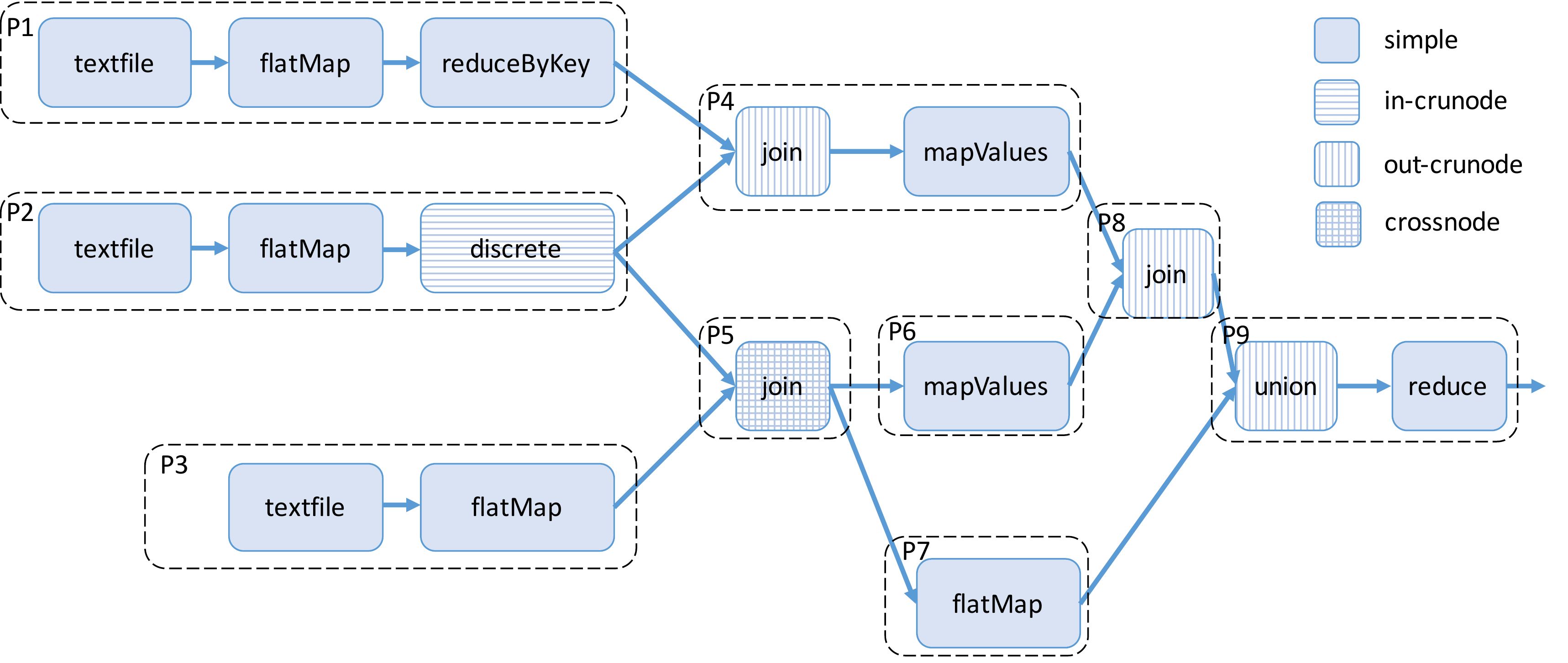}\vspace{-3mm}
\caption{\scriptsize \textbf{Job DAG example}. An example of a parallel job represented as a DAG. Each node corresponds to an operation resulting RDD that can be executed over a parallel cluster (e.g., a map, reduce, or join operation). DAG edges indicate precedence. 
Simple, crunodes (in/out) and cross nodes are represented with solid or lined textures.}\vspace*{-1.3em}
\label{FIG:DAG}
\end{figure}

Caching an RDD resulting from a computation step in a job like the one appearing in Fig.~\ref{FIG:DAG} can have significant computational benefits when jobs may exhibit \emph{computational overlaps}: not only are jobs executed over the same data, but also consist of operations that are repeated across multiple jobs. This is illustrated in Fig.~\ref{FIG:JOBARRIVALS}: jobs may be distinct, as they comprise different sets of operations, but certain subsets of operations (shown as identically colored subgraphs in the DAG of Fig.~\ref{FIG:JOBARRIVALS}) are (a) the same, i.e., execute the same primitives (maps, joins, etc.) and (b) operate over the same data. 

\vspace{-2mm}

Computational overlaps arise in practice for two reasons. The first is that operations performed by companies are often applied to the same data-pipeline: companies collect data generated by their applications and users, which they maintain in the cloud, either directly on a distributed file system like HDFS, or on NoSQL databases (like Google's Datastore~\cite{barrett2008under} or Apache HBase~\cite{hbase}). Operations are therefore performed on the same source of information: the latest data collected within a recent period of time.
The second reason for computational overlaps is the abundance of commonalities among computational tasks in data-parallel processing. Commonalities occur in several classic data-mining and machine learning operations heavily utilized in inference and prediction tasks (such as predictions of clickthrough rates and user profiling). 
We give some illustrative examples below:



\vspace{-2mm}
\noindent\textbf{Projection and Selection.} The simplest common preprocessing steps are \emph{projection} and \emph{selection}~\cite{maier1983theory}. For example, computing the mean of a variable $\mathtt{age}$ among tuples satisfying the predicate $\mathtt{gender}=\mathtt{female}$ and $\mathtt{gender}=\mathtt{female}\land \mathtt{income}\geq 50$K might both first reduce a dataset by selecting rows in which $\mathtt{gender}=\mathtt{female}$. Even in the absence of a relational database, as in the settings we study here, projection (i.e., maintaining only certain feature columns) and selection (i.e., maintaining only rows that satisfy a predicate) are common. For example, building a classifier that predicts whether a user would click on an advertisement relies upon first restricting a dataset containing all users to the history of the user's past clicking behavior. This is the same irrespective of the advertisement for which the classifier is trained. 

%

\vspace{-2mm}
\noindent\textbf{Supervised Learning.} Supervised learning tasks such as regression and classification~\cite{trevor2001elements}, i.e., training a model from features for the purpose of predicting a label (e.g., whether a user will click on an advertisement or image) often involve common operations that are label-independent. For example, performing ridge regression first requires computing the co-variance of the features~\cite{trevor2001elements}, an identical task irrespective of the label to be regressed. Similarly, kernel-based methods like support vector machines require precomputing a kernel function across points, a task that again remains the same irrespective of the labels to be regressed~\cite{scholkopf2001learning}. Using either method to, e.g., regress the click-through rate of an ad, would involve the same preprocessing steps, irrespectively of the labels (i.e., clicks pertaining to a specific ad) being regressed.

\vspace{-2mm}
\noindent\textbf{Dimensionality Reduction.} Preprocessing also appears in the form of \emph{dimensionality reduction}: this is a common preprocessing step in a broad array of machine learning and data mining tasks, including regression, classification, and clustering. Prior to any such tasks, data is first projected in a lower dimensional space that preserves, e.g., data distances. There are several approaches to doing this, including principal component analysis~\cite{jolliffe2002principal}, compressive sensing~\cite{eldar2012compressed}, and training autoregressive neural networks~\cite{gregor2014deep}, to name a few. In all these examples, the same projection would be performed on the data prior to subsequent processing, and be reused in the different tasks described above.

\vspace{-2mm}
To sum up, the presence of computational overlaps across jobs gives rise to a tremendous opportunity of reducing computational costs. Such overlaps can be exploited precisely through the caching functionality of a data-parallel framework. 
If a node in a job is cached (i.e., results are \emph{memoized}), then neither itself nor any of its predecessors need to be recomputed. 

\vspace{-3mm}
\subsection{Problems and Challenges}
\vspace{-2mm}

Designing caching schemes poses several significant challenges. 
To begin with, making caching decisions is an inherently combinatorial problem. Given (a) a storage capacity constraint, (b) a set of jobs to be executed, (c) the size of each computation result, and (d) a simple linear utility on each job, the problem is reduced to a knapsack problem, which is NP-hard. The more general objectives we discussed above also lead to NP-hard optimization problems~\cite{fleischer2011tight}. 
Beyond this inherent problem complexity, even if jobs are selected from a pool of known jobs (e.g., classification, regression, querying), the sequence 
to submit jobs within a given time interval \emph{is a priori unknown}. The same may be true about statistics about upcoming jobs, such as the frequency with which they are requested.
%
%
To that end, a practical caching algorithm must operate in an \emph{adaptive} fashion: it needs to make online decisions on what to cache as new jobs arrive, and adapt to changes in job frequencies.

\vspace{-1mm}
In Spark, LRU is the default policy for evicting RDDs when the cache is full. There are some other conventional caching algorithms such as LRU variant~\cite{LRU-K} that maintains the most recent accessed data for future reuse, and ARC~\cite{NM-ARC} and LRFU~\cite{LRFU} that consider both frequency and recency in the eviction decisions. 
When the objective is to minimize total work, these conventional caching algorithms are woefully inadequate, leading to arbitrarily suboptimal caching decisions~\cite{ioannidis2016adaptive}. Recently, a heuristic policy~\cite{geng2017lcs}, named ``Least Cost Strategy'' (LCS), was proposed to make eviction decisions based on the recovery temporal costs of RDDs. However, this is a heuristic approach and again comes with no guarantees.
In contrast, we intend to leverage Spark's internal caching mechanism to implement our caching algorithms and deploy and evaluate them over the Spark platform, while also attaining formal guarantees.

\vspace{-3mm}
\section{Algorithm Design}
\label{SEC:AD}
\vspace{-2mm}

In this section, we introduce a formal mathematical model for making caching decisions that minimize the expected total work, i.e., the total expected computational cost for completing all jobs.  
The corresponding caching problem is NP-hard, even in an offline setting where the popularity of jobs submitted to the cluster is  \emph{a priori known}. 
Nevertheless, we show it is possible to pose this optimization problem as a submodular maximization problem subject to knapsack constraints. This allows us to produce a $1-1/e$ approximation algorithm for its solution. Crucially, when job popularity is \emph{not known}, we have devised an adaptive algorithm for determining caching decisions probabilistically, that makes caching decisions lie within $1-1/e$ approximation from the offline optimal, in expectation. 

\vspace{-2mm}
\subsection{DAG/Job Terminology}
\vspace{-2mm}

We first introduce the terminology  we use in describing caching algorithms.
Consider a job represented as a DAG as shown in Fig.~\ref{FIG:DAG}. Let $G(V,E)$ be the graph representing this DAG, whose nodes are denoted by $V$ and edges are denoted by $E$. Each node is associated with an operation to be performed on its inputs (e.g., map, reduce, join, etc.). 
These operations come from a well-defined set of operation primitives (e.g., the operations defined in Spark). For each node $v$, we denote as $\op(v)$ the operation that $v\in V$ represents. The DAG $G$ as well as the labels $\{\op(v),v\in V\}$ fully determine the job.
A node $v\in V$ is a \emph{source} if it contains no incoming edges, and a \emph{sink} if it contains no outgoing edges. Source nodes naturally correspond to operations performed on ``inputs'' of a job (e.g., reading a file from the hard disk), while sinks correspond to ``outputs''. 
Given two nodes $u,v\in V$, we say that $u$ is a \emph{parent} of $v$, and that $v$ is a \emph{child} of $u$, if $(u,v)\in E$. We similarly define \emph{predecessor} and \emph{successor} as the transitive closures of these relationships. 
For $v\in V$, we denote by $\prd(v)\subset V$, $\scc(c)\subset V$ the sets of predecessors and successors of $v$, respectively.
Note that the parent/child relationship is the opposite to usually encountered in trees, where edges are usually thought of as pointing away from the root/sink towards the leaves/sources.
%
%
We call a DAG a \emph{directed tree} if (a) it contains a unique sink, and (b) its undirected version (i.e., ignoring directions) is acyclic. 

\vspace{-2mm}
\subsection{Mathematical Model}
\vspace{-2mm}

Consider a setting in which all jobs are applied to the same dataset; this is without loss of generality, as multiple datasets can be represented as a single dataset--namely, their union--and subsequently adding appropriate projection or selection operations as preprocessing to each job. Assume further that each DAG is a directed tree. Under these assumptions, let $\jobset$ be the set of all possible jobs that can operate on the dataset. We assume that jobs $G\in\jobset$ arrive according to a stochastic stationary process with rate $\lambda_G>0$. Recall that each job $G(V,E)$ comprises a set of nodes $V$, and that each node $v\in V$ corresponds to an operation $\op(v)$. We denote by as $c_v\in \reals_+$ the time that it takes to execute this operation given the outputs of its parents, and $s_v\in \reals_+$ be the size of the output of $\op(v)$, e.g., in Kbytes. Without caching, the \emph{total-work} of a job $G$ is then given by
$W(G(V,E)) = \sum_{v\in V} c_v.$
We define the \emph{expected total work} as: 
\begin{align}
\bar{W} =\sum_{G\in \mathcal{G}} \lambda_G \cdot W(G) = \sum_{G(V,E)\in \mathcal{G}} \lambda_{G(V,E)} \sum_{v\in V} c_v.
\end{align}
%
We say that two nodes $u,u'$ are \emph{identical}, and write $u=u'$, if both these nodes and all their predecessors involve exactly the same operations. 
%
%
We denote by $\nodeset$ 
the union of all nodes of DAGs in $\jobset$. 
A \emph{caching strategy} is a vector $x=[x_v]_{v\in\nodeset}\in \{0,1\}^{|\nodeset|}$, where $x_v\in\{0,1\}$ is a binary variable indicating whether we have cached the outcome of node $v$ or not.
As jobs in \jobset{} are directed trees, when node $v$ is cached,  \emph{there is no need to compute that node or any predecessor of that node}. 
Hence, under a caching strategy $x$, the total work of a job $G$ becomes:
%
\begin{align}
\textstyle W =\sum_{v\in V} c_v (1-x_v)\prod_{u\in \scc(v)} (1-x_u).
\end{align}
%
Intuitively, this states that the cost $c_v$ of computing $\op(v)$ needs to be paid if and only if \emph{neither} $v$ \emph{nor} any of its successors have been cached.

\vspace{-3mm}
\subsection{Maximizing the Caching Gain: Offline Optimization} 

Given a cache of size $K$ Kbytes, we aim to solve the following optimization problem:

\vspace{-2mm}
\begin{subequations}\label{maxcachegain}
\small{{\hspace*{\stretch{1}} \textsc{MaxCachingGain}\hspace{\stretch{1}} }}
\begin{align}
\text{Max:}& & & F(x) \!=\! \bar{W} \!-\!\! \sum_{G\in\jobset}\!\!\lambda_G W(G,x)\!\label{obj}\\
\text{}&&& =\!\!\! \sum_{G(V,E)\in\jobset}\!\!\!\!\!\! \lambda_G\!\sum_{v\in V}\!c_v\big[1- (1\!-\!x_v)\!\!\!\!\!\prod_{u\in \scc(v)}\!\!\! (1\!-\!x_u)\big] \\
\text{Sub.~to:}&&& \textstyle\sum_{v\in \nodeset} s_vx_v\leq K, \quad
x_v\in\{0,1\}, \text{ for all } v\in \nodeset.\label{intcont}
\end{align}
\end{subequations}

\vspace{-2mm}
Following~\cite{ioannidis2016adaptive}, we call function $F(x)$ the \emph{caching gain}: this is the reduction on total work due to caching. This offline problem
is NP-hard~\cite{shanmugam2013femtocaching}. 
Seen as an objective over the set of nodes $v\in \nodeset$ cached, $F$ is a \emph{monotone, submodular} function. Hence, \eqref{maxcachegain} is a submodular maximization problem with a knapsack constraint. When all outputs have the same size, the classic greedy algorithm by Nemhauser et al.~\cite{nemhauser} yields a $1-1/e$ approximation. In the case of general knapsack constraints, there exist well-known modifications of the greedy algorithm that yields the same approximation ratio~\cite{sviridenko-submodular,krause-submodular,kulik2009maximizing}.

\vspace{-2mm}
Beyond the above generic approximation algorithms for maximizing submodular functions, \eqref{maxcachegain} can be solved by \emph{pipage rounding}~\cite{ageev2004pipage}.
In particular, there exists a concave function $L:[0,1]^{|\nodeset|}$ such that: 
\begin{align}(1-1/e)L(x)\leq F(x) \leq L(x), \quad\text{for all}~x\in[0,1]^{|\nodeset|}\label{sandwitch}.\end{align}
This \emph{concave relaxation} of $F$ is given by:
\begin{align}\label{relaxation}
L(x) = \sum_{G(V,E)\in\jobset} \lambda_G\sum_{v\in V}c_v\min\big\{1, x_v+\sum_{u\in \scc(v)}x_u\big\}.
\end{align}
Pipage rounding solves \eqref{maxcachegain} by replacing objective $F(x)$ with its concave approximation $L(x)$ and relaxing the integrality constraints \eqref{intcont} to the convex constraints $x\in[0,1]^{|\nodeset|}$. The resulting optimization problem is convex--in fact, it can be reduced to a linear program, and thus solved in linear time. Having solved this convex optimization problem, the resulting fractional solution is subsequently rounded to produce an integral solution. Several polynomial time rounding algorithms exist (see, e.g., \cite{ageev2004pipage,swaprounding}, and   \cite{kulik2009maximizing} for knapsack constraints). Due to \eqref{sandwitch} and the specific design of the rounding scheme, the resulting integral solution is guaranteed to be within a constant approximation of the optimal \cite{ageev2004pipage,kulik2009maximizing}.

\vspace{-2mm}
\subsection{An Adaptive Algorithm with Optimality Guarantees}
\label{subsec:adaptive}

\begin{figure*}[ht]
\centering
\includegraphics[width=0.84\textwidth]{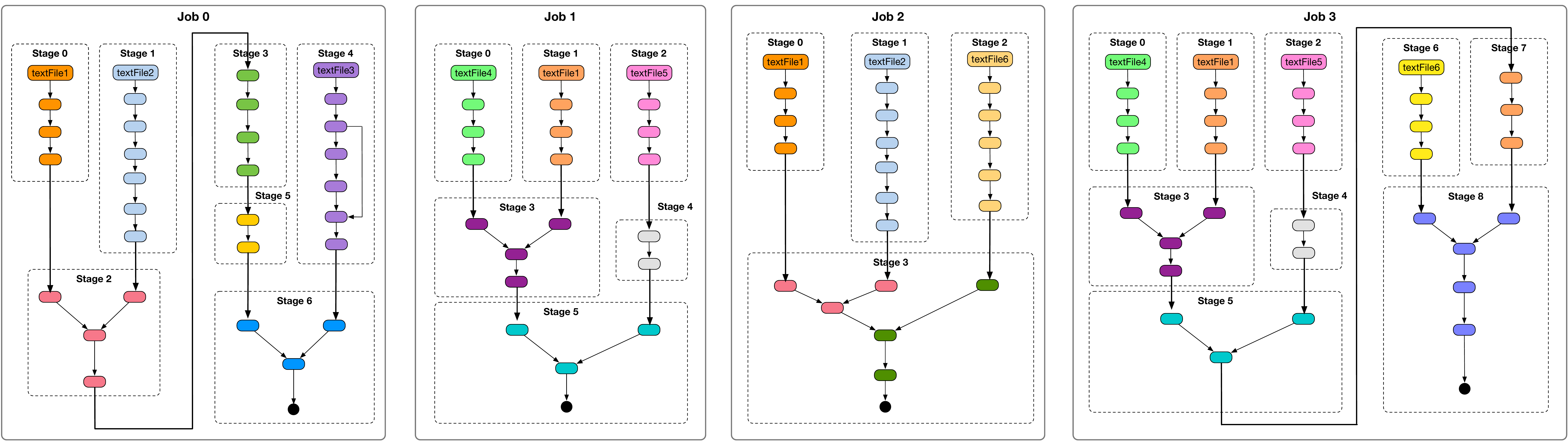}\vspace{-3mm}
\caption{\scriptsize {\bf{An example of RDD dependency in synthetic jobs.}} Denote $J_x.S_y$ as stage $y$ in job $x$, then we have $J_0.S_0 = J_1.S_1=J_2.S_0=J_3.S_1$, 
$J_1.S_{0\sim 5}=J_3.S_{0\sim 5}$, and 
$J_0.S_{0\sim 1}=J_2.S_{0\sim 1}$. 
Unfortunately, even sharing the same computational overlap, by default these subgraphs will be assigned with different stage/RDD IDs by Spark since they are from different jobs.
\vspace*{-1em}
}
\label{FIG:EXP-DAG}
\end{figure*}


\begin{algorithm}[t]
\scriptsize
\SetAlFnt{\footnotesize}
\SetKwFunction{FuncIterateJobs}{processJobs}
\SetKwProg{ProcIterateJobs}{Procedure}{}{}

\SetKwFunction{FuncIterate}{processJob}
\SetKwProg{ProcIterate}{Procedure}{}{}

\SetKwFunction{FuncCalCost}{estimateCost}
\SetKwProg{ProcCalCost}{Procedure}{}{}

\SetKwFunction{FuncSaveRDD}{updateCache}
\SetKwProg{ProcSaveRDD}{Procedure}{}{}

\SetKwIF{If}{ElseIf}{Else}{if}{then}{else if}{else}{endif}

\ProcIterateJobs{\FuncIterateJobs{\jobset}}
{   
    $C_{\jobset}$ = Historical RDD access record\;
    $C_G$ = Current job RDD access record\;
    \For{$G\in\jobset$}
    {
        
        processJob($G(V,E)$, $C_G$)\;
        updateCache($C_G$, $C_{\jobset}$)\;
    }
}    
\ProcIterate{\FuncIterate{$G(V,E)$, $C$}}
{   
    $C_G$.clear()\;
       \For{v$\in$V}
    {
        v.accessed=False\;
        toAccess=set(DAG.sink())\;
        \While{toAccess$\neq \emptyset $}
        {
            v=toAccess.pop()\; 
            $C_G$[v]=estimateCost(v)\;

            \If{not v.cached}
            {
                \For{u $\in$ v.parents}
                {
                    \If{not u.accessed}
                    {
                        toAccess.add(u)\;
                    }
                }
            }
            access(v); /* Iterate RDD $v$. */ \\
            v.accessed=True\;
        }
    }
    \KwRet\;
}            
\ProcCalCost{\FuncCalCost{v}}
{
    cost=compCost[v]; /* If all parents are ready. */\\
    toCompute=v.parents /* Check each parent. */\\
    \While{toCompute $\neq \emptyset$}
    {
        u=toCompute.pop()\;
        \If {not (u.cached or u.accessed or u.accessedInEstCost)}
        {
            cost+=compCost[u]\;
            toCompute.appendList(u.parents)\;    
            u.accessedInEstCost=True\; 
        }
    }
    \KwRet cost;
}
\ProcSaveRDD{\FuncSaveRDD{$C_G$, $C_{\jobset}$}}
{
    \For {$v \in C_{\jobset}$}
    {
        \If{$v \in C_G$}
        {
            $C_{\jobset}[v]  =(1-\beta)\times C_{\jobset}[v] + \beta \times C_G[v]$\;
        }
        \Else
        {
            $C_{\jobset}[v]  =(1-\beta)\times C_{\jobset}[v] $\;
        }
        updateCacheByScore($C_{\jobset}$)\;
    }
    \KwRet\;
}
\caption{\small A Heuristic Caching Algorithm.}
\label{ALG:1}
\end{algorithm}

\vspace{-2mm}
As discussed above, if the arrival rates $\lambda_G$, $G\in\jobset$, are known, we can determine a caching policy within a constant approximation from the optimal solution to the (offline) problem \textsc{MaxCachingGain} by solving a convex optimization problem. 
In practice, however, the arrival rates $\lambda_G$ may \emph{not} be known. To that end, we are interested in an \emph{adaptive} algorithm, that converges to caching decisions \emph{without any prior knowledge of job arrival rates} $\lambda_G$. 
Building on \cite{ioannidis2016adaptive}, we propose an adaptive algorithm for precisely this purpose. We describe the details of this adaptive algorithm in {\intechreport{the Appendix~\ref{app:overview}.}{our technical report \cite{techrep}.}}
In short, our adaptive algorithm performs \emph{projected gradient ascent} over concave function $L$, given by \eqref{relaxation}. 
That is, our algorithm maintains at each time a fractional $y\in[0,1]^{|\nodeset|}$, capturing the probability with which each RDD should be placed in the cache. Our algorithm collects information from executed jobs; this information is used to produce an estimate of the gradient $\nabla L(y)$. In turn, this is used to adapt the probabilities $y$ that we store different outcomes. Based on these adapted probabilities, we construct a randomized placement $x$ satisfying the capacity constraint \eqref{intcont}. We can then show that the resulting randomized placement has the following property: 
\begin{thm} \label{mainthm}If $x(t)$ is the placement at time $t$, then 
$\textstyle\lim_{t\to \infty} \mathbb{E}[F(x(t))] \geq \big(1 -{1}/{e}\big) F(x^*),$
where $x^*$ is an optimal solution to the offline problem \textsc{MaxCachingGain} (Eq.~\eqref{maxcachegain}).
\end{thm}
The proof of Thm.~\ref{mainthm} can be found in  \intechreport{Appendix~\ref{app:proofofmainthm}.}{our technical report~\cite{techrep}.}

\subsection{A Heuristic Adaptive Algorithm}

Beyond attaining such guarantees, our adaptive algorithm gives us a great intuition to prioritize computational outcomes. Indeed, the algorithm prioritizes nodes $v$ that have a high gradient component $\partial L/\partial x_v$ and a low size $s_v$. Given a present placement, RDDs should enter the cache if they have a high value w.r.t.~the following quantity \intechreport{(Appendix~\ref{app:proofofmainthm})}{~\cite{techrep}}:
\begin{align}\textstyle \frac{\partial L}{\partial x_v}/s_v \simeq \left(\textstyle\sum_{G\in \jobset: v\in G}\lambda_G \times \Delta(w)\right)/s_v, \label{approx}\end{align}
%
%
where $\Delta(w)$ is the difference in total work if $v$  is not cached.
This intuition is invaluable in coming up with useful heuristic algorithms for determining what to place in a cache. In contrast to, e.g., LRU and LFU, that prioritize jobs with high request rate, Eq.~\eqref{approx} suggests that a computation should be cached if (a) it is requested often, (b) caching it can lead to a significant reduction on the total work, and (c) it has small size. Note that (b) is \emph{dependent on other caching decisions made by our algorithm}. Observations (a), (b), and (c) are intuitive, and the specific product form in \eqref{approx} is directly motivated and justified by our formal analysis.
They give rise to the following simple heuristic adaptive algorithm: for each job submitted, maintain a moving average of  (a) the request rate of individual nodes it comprises, and (b) the cost that one would experience if these nodes are not cached. Then, place in the cache only jobs that have a high such value, when scaled by the size $s_v$. 

\vspace{-2mm}
Alg.~\ref{ALG:1} shows the main steps of our heuristic adaptive algorithm. 
It updates the cache (i.e., storage memory pool) after the execution of each job (line 5) based on decisions made in the $updateCache$ function (line 6), which considers both the historical (i.e., $C_{\jobset}$) and current RDD (i.e., $C_G$) cost scores.
Particularly, when iterating RDDs in each job following a recursive fashion, 
an auxiliary function $estimateCost$ is called to calculate and record the temporal and spatial cost of each RDD in that job (see line 14 and lines 22 to 31).
Notice that $estimateCost$ does not actually access any RDDs, but conducts DAG-level analysis for cost estimation which will be used to determine cache contents in the $updateCache$ function.
In addition, a hash mapping table is also used to record and detect computational overlap cross jobs (details see in our implementation in Sec.~\ref{SUBSEC:PE-SI}). 
After that, we iterate over each RDD's parent(s) (lines 16 to 18). Once all its parent(s) is(are) ready, we access (i.e., compute) the RDD (line 19). 
Lastly, the $updateCache$ function first updates the costs of all accessed RDDs to decide the quantities cost collected above with a moving average window using a decay rate of $\beta$, implementing an Exponentially Weighted Moving Average (EWMA). 
Next, $updateCache$  makes cross-job cache decisions based on the sorting results of the moving average window by calling the $updateCacheByScore$ function. 
The implementation of this function can (1) refresh the entire RAM by top score RDDs; or (2)  evict lower score old RDDs to insert higher score new RDDs. 
%


\vspace{-3mm}
\section{Performance Evaluation}
\label{SEC:PE}
\vspace{-2mm}

\begin{figure*}[ht]
\centering
\includegraphics[width=0.85\textwidth]{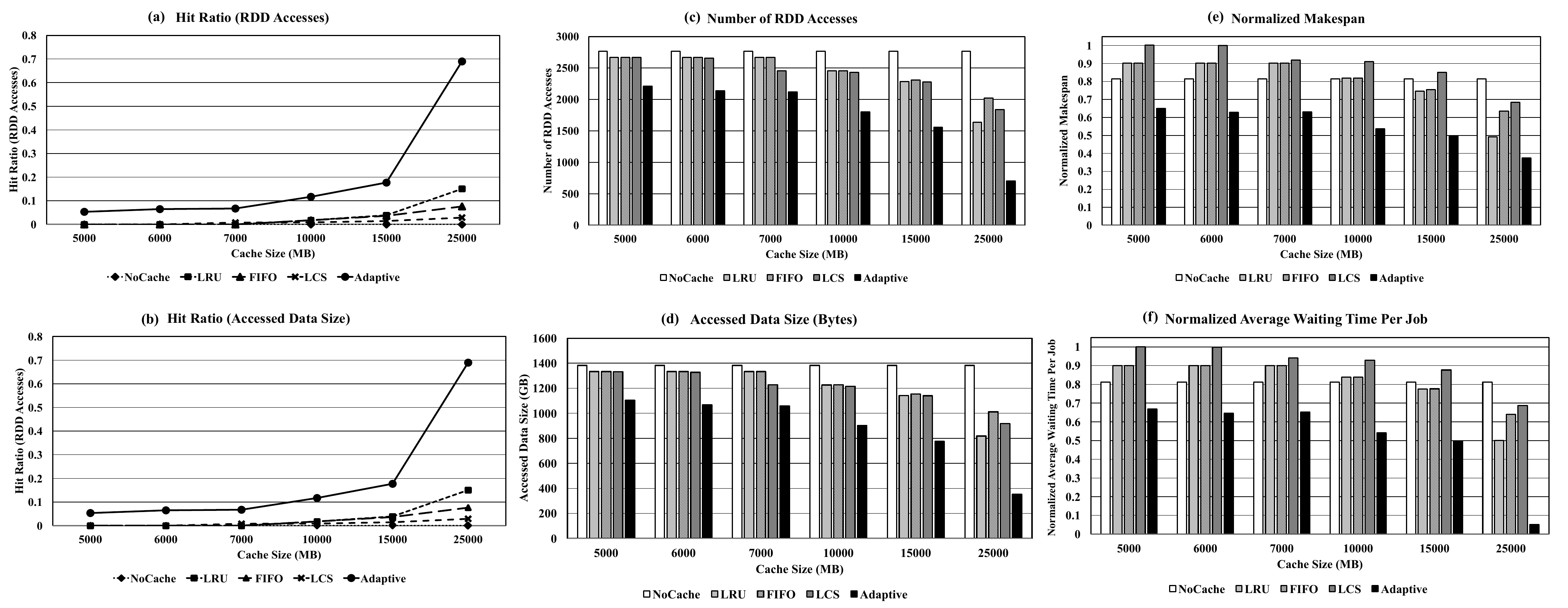}\vspace{-3mm}
\caption{\small Hit ratio, access number and total work makespan results of large scale simulation experiments.
\vspace*{-2.3em}
}
\label{FIG:EXP-SIM}
\end{figure*}

In this section, we first demonstrate the performance of our adaptive caching algorithm ( Alg.~\ref{ALG:1})  on a simple illustrative example.  
We then build a simulator to analyze the  performance of large-scale synthetic traces with complex DAGs. 
Lastly, we validate the effectiveness of our adaptive  algorithm  by conducting real experiments in Apache Spark with real-world machine learning workloads.


\vspace{-3mm}
\subsection{Numerical Analysis}
\label{SUBSEC:PE-NA}
\vspace{-2mm}

We use a simple example to illustrate how our adaptive algorithm (i.e., Alg.~\ref{ALG:1}) performs w.r.t minimizing total work. 
This example is specifically designed to illustrate that our  algorithm  significantly outperforms the default LRU policy used in Spark. 
Assume that we have 5 jobs ($J_0$ to $J_4$) each consisting of 3 RDDs, the first 2 of which are common across jobs. 
That is, $J_0$'s DAG is $R_0$$\rightarrow$$R_1$$\rightarrow$$R_2$, 
$J_1$ is $R_0$$\rightarrow$$R_1$$\rightarrow$$R_3$, 
$J_2$ is $R_0$$\rightarrow$$R_1$$\rightarrow$$R_4$, etc. 
The calculation time for $R_1$ is 100 seconds while the calculation time for other RDDs (e.g., $R_2$, $R_3$,...) is 10 seconds. 
We  submit this sequence of jobs twice, with the interarrival time of 10 seconds between jobs. Thus, we have 10 jobs in a sequence of \{$J_0$, $J_1$, ..., $J_4$, $J_0$, $J_1$, ..., $J_4$\}. We set the size of each RDD as  500MB and the cache capacity as 500MB as well. Hence,  at most one RDD can be cached at any  moment.

Table~\ref{TAB:sample} shows the experimental results of this simple example under LRU and our algorithm. Obviously, LRU cannot well utilize the cache because the recently cached RDD (e.g., $R_2$) is always evicted by the newly accessed RDD (e.g., $R_3$). As a result, none of the RDDs are hit under the LRU policy. By producing an estimation of the gradient on RDD computation costs, our algorithm instead places $R_1$ in the cache after the second job finishes and thus achieves a higher hit ratio of 36\%, i.e., 8 out of 22 RDDs are hit. Total work (i.e., the total calculation time for finishing all jobs) is significantly reduced as well under our algorithm.

\begin{table}[h]
\scriptsize
\vspace{-0.1in}
\caption{Caching results of the simple case.}
\label{TAB:sample}
\centering
\begin{tabular}{|p{10mm}|p{2.5mm}|p{2.5mm}|p{2.5mm}|p{2.5mm}|p{2.5mm}|p{2.5mm}|p{9mm}|p{13mm}|}
\hline
Policy & $J_0$ & $J_1$ & $J_2$ & $J_3$ & ... & $J_4$ & hitRatio & totalWork \\ \hline 
LRU & $R_2$ & $R_3$ & $R_4$ & $R_5$ & ... & $R_6$ &0.0\% & 1100 \\ \hline
Adaptive & $R_2$ & $R_1$ & $R_1$ & $R_1$ & ... & $R_1$ &36.4\% & 300 \\ \hline
\end{tabular}
\vspace{-0.1in}
\end{table}



\vspace{-2mm}
\subsection{Simulation Analysis}
\label{SUBSEC:PE-SA}
\vspace{-2mm}

To further validate the effectiveness of our proposed algorithm, we scale up our synthetic trace by randomly generating a sequence of 1000 jobs to represent real data analysis applications with complex DAGs. Fig.~\ref{FIG:EXP-DAG} shows an example of some jobs' DAGs from our synthetic trace, where some jobs  include stages and RDDs with the same generating logic chain. For example,  stage 0 in $J_0$ and stage 1 in $J_1$ are identical, but their RDD IDs are different and will be computed twice. 
%
%
On average, each of these jobs consists of six stages and each stage has six RDDs. 
The average RDD size is 50MB. 
We use a decay rate of $\beta=0.6$. 


\vspace{-2mm}
We implement four caching algorithms for comparison: (1) NoCache: a baseline policy, which forces Spark to ignore all user-defined {\em cache}/{\em persist} demands, and thus provides the lower bound of caching performance; (2) LRU: the default policy used in Spark, which evicts the least recent used RDDs; (3) FIFO: a traditional policy which evicts the earliest RDD in the RAM; and (4) LCS: a recently proposed policy, called ``Least Cost Strategy''~\cite{geng2017lcs}, which uses a heuristic  approach to calculate each RDD's recovery temporal cost to make eviction decisions. 
%
%
%
The main metrics include 
(a) {\em RDD hit ratio} that is calculated as  the ratio between the number of RDDs hit in the cache and the total number of accessed RDDs, or the ratio between the size of RDDs hit in the cache and the total size of accessed RDDs;
(b) {\em Number of accessed RDDs} and {\em total amount of accessed RDD data size} that need to be accessed through the experiment; 
(c) {\em Total work} (i.e., makespan) that is the total calculation time for finishing all jobs;  
and 
(d) {\em Average waiting time} for each job.
%
%
%

Fig.~\ref{FIG:EXP-SIM} depicts the performance of the five caching algorithms.
We conduct a set of simulation experiments by configuring different cache sizes for storing RDDs. 
Clearly, our  algorithm (``Adaptive") significantly improves the hit ratio (up to 70\%) across different cache sizes, as seen Fig.~\ref{FIG:EXP-SIM}(a) and (b). 
In contrast, the other  algorithms start to hit RDDs (with hit ratio up to 17\%) only when the cache capacity becomes large. 
Consequently, our proposed algorithm reduces the number of RDDs that need to be accessed and calculated (see Fig.~\ref{FIG:EXP-SIM}(c) and (d)), which further saves the overall computation costs, i.e., the total work in Fig.~\ref{FIG:EXP-SIM}(e) and (f). 
We also notice that such an improvement from ``Adaptive" becomes more significant when we have a larger cache space for RDDs, which  indicates that our adaptive algorithm is able to better detect and utilize those shareable and reusable RDDs across jobs.  


\vspace{-3mm}
\subsection{Spark Implementation}
\label{SUBSEC:PE-SI}
\vspace{-2mm}

\begin{figure}[ht]
\centering
\includegraphics[width=0.31\textwidth]{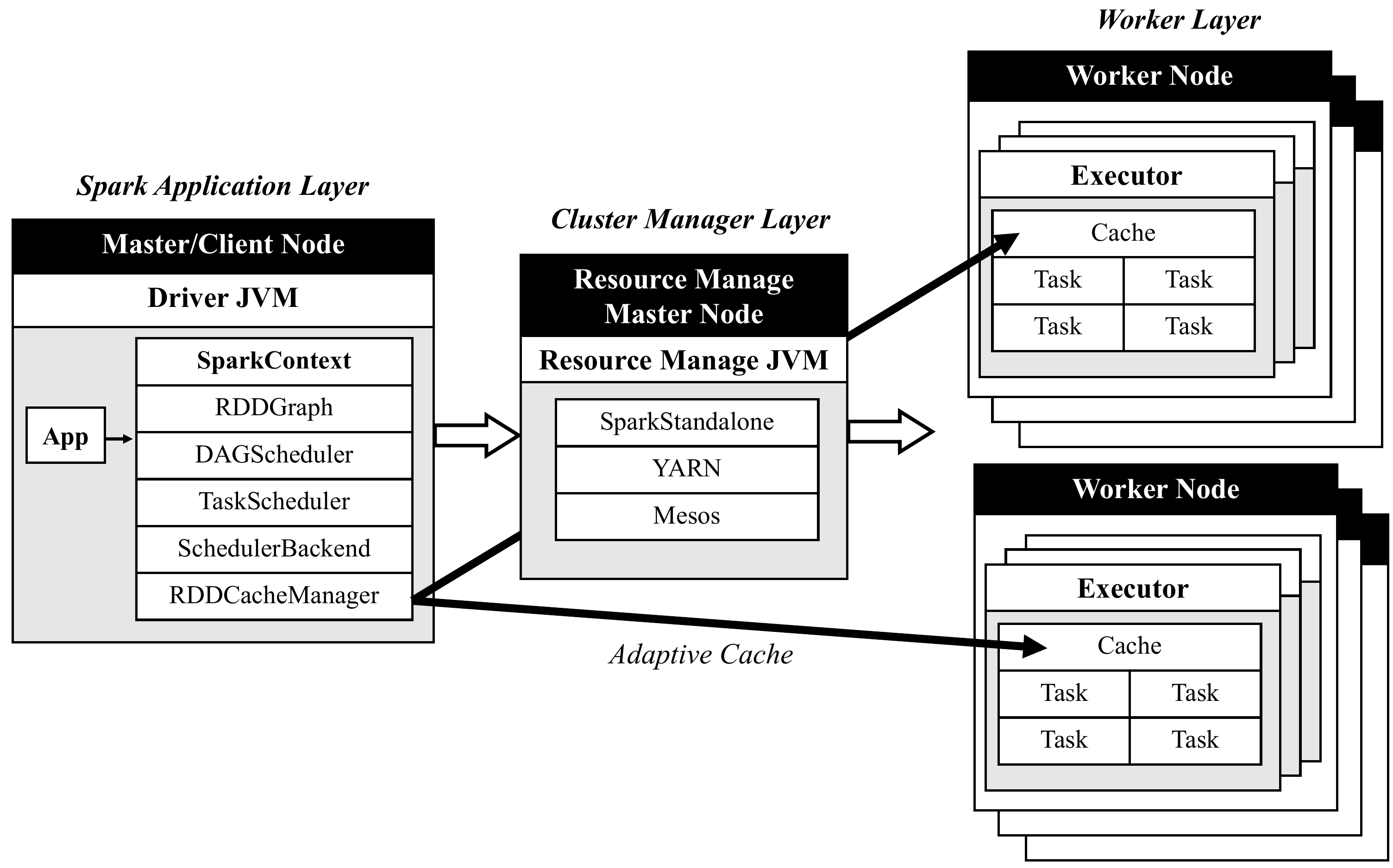}\vspace{-3mm}
\caption{\small Module structure view of our Spark implementation, where our proposed {\em RDDCacheManager} module cooperates with {\em cache} module inside each worker node.
}
\label{FIG:EXP-SP-ARC}
\end{figure}

We further evaluate our cache algorithm  by integrating our methodology into  Apache Spark 2.2.1, hypervised by VMware Workstation 12.5.0. 
Table~\ref{TAB:EV-SPEC} summarizes the details of our testbed configuration. 
%
In Spark, the memory space is divided into four pools: storage memory, execution memory, unmanaged memory and reserved memory. 
Only storage and execution memory pools (i.e., $UnifiedMemoryManager$) are used to store runtime data of Spark applications. 
Our implementation focuses on  storage memory,  
which stores cached data (RDDs), internal data propagated through the cluster, and temporarily unrolled serialized data.
Fig.~\ref{FIG:EXP-SP-ARC} further illustrates the main architecture of modules in our implementation. 
In detail, different from Spark's built-in caching that responds to {\em persist} and {\em unpersist} APIs, we build an {\em RDDCacheManager} module in the {\em Spark Application Layer} to communicate with cache modules in the {\em Worker Layer}. 
Our proposed module maintains statistical records (e.g., historical access, computation  overhead, DAG dependency, etc.), and automatically  decides which new RDDs to be cached and which existing RDDs to be evicted when the cache space is full. 

\vspace{-1mm}
\small
\begin{table}[th]
    \center
    \caption{Testbed configuration.}    
    \label{TAB:EV-SPEC}
    \begin{tabular}{|c|c|}
        \hline
        \textbf{Component} & \textbf{Specs} \\ \hline 
        Host Server & Dell PowerEdge T310  \\ \hline
        Host Processor &  Intel Xeon CPU  X3470   \\ \hline
        Host Processor Speed & 2.93GHz \\ \hline
        Host Processor Cores & 8 Cores \\ \hline
        Host Memory Capacity & 16GB DIMM DDR3  \\ \hline
        Host Memory Data Rate & 1333 MHz \\ \hline
        Host Hypervisor & VMware Workstation 12.5.0 \\ \hline
        Big Data Platform & Apache Spark 2.2.1 \\ \hline
        Storage Device &     Western Digital WD20EURS  \\ \hline
        Disk Size & 2TB \\ \hline
        Disk Bandwidth & SATA 3.0Gbps \\ \hline
        Memory Size Per Node & 1 GB \\ \hline
        Disk Size Per Node & 50 GB\\ \hline

    \end{tabular}    
    \end{table}
\normalsize
\vspace{-1mm}

%
\begin{figure}[h]
\centering
\includegraphics[width=0.29\textwidth]{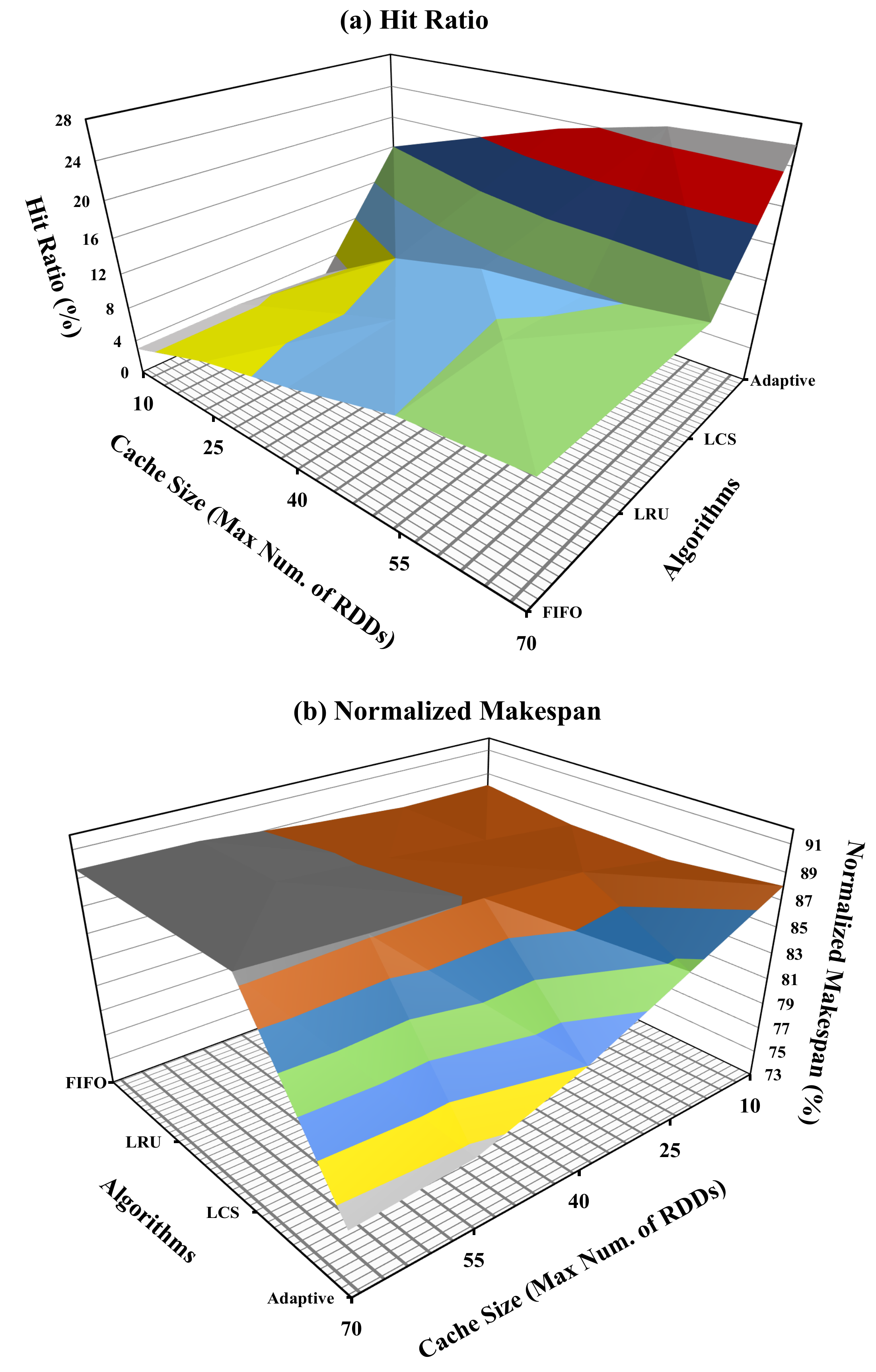}\vspace{-3mm}
\caption{\small Hit ratio and normalized makespan results of a stress testing on cache-unfriendly {\em Ridge Regression} benchmark with different cache sizes under four cache algorithms.\vspace*{-1.3em}}
\label{FIG:EXP-SP-REC}
\end{figure}

We select  {\em Ridge Regression}~\cite{hoerl1970ridge} as a benchmark because it is a ubiquitous technique,  widely applied in machine learning and data mining applications~\cite{li2013enhanced, huang2012extreme}.
%
%
The input database we use is a huge table containing thousands of entries (i.e., rows), and each entry has more than ten features (i.e., columns).  
More than hundred Spark jobs are repeatedly generated with an exponential arrival rate. 
Each job's DAG contains at least one {\em Ridge Regression}-related subgraph, which regresses a randomly selected feature column (i.e., target) by a randomly selected subset of the remaining feature columns (i.e., source), i.e., $f_t=\Re (\vec {f_s})$, where $f_t$ is the target feature, and $\Re(\vec {f_s})$ is the regressed correlation function with an input of source feature vector $\vec {f_s}$.
Moreover, different jobs may share the same selections of target and source features, and thus they may have some RDDs with exactly the same generating logic chain (i.e., a subset of DAGs). 
Unfortunately, the default Spark cannot identify RDDs with the same generating logic chain if they are in {\em different} jobs.
In order to identify these reusable and identical RDDs, our proposed {\em RDDCacheManager} uses a mapping table to records each RDD's generating logic chain {\em across} jobs (by importing our customized header files into the benchmark), i.e., we denote $RDD_x$ by a hashing function $key\leftarrow hash(G_x(V,E) )$, where $G_x(V,E)$ is the subgraph of $RDD_x$ ($V$ is the set of all ancestor RDDs and $E$ is the set of all operations along the subgraph). 
Since not all operations are deterministic~\cite{determ} (e.g., {\em shuffle} operation on the same input data may result in different  RDDs),  we only monitor those deterministic operations which  guarantee the same output under the same input. 




\vspace{-3mm}
Rather than scrutinizing the cache-friendly case where our adaptive algorithm appears to work well as shown in Sec.~\ref{SUBSEC:PE-SA}, 
it will be more interesting to study the performance under the cache-unfriendly case (also called ``stress test''~\cite{wagner2005stresstest}), where the space size of different combinations of source and target features is comparatively large, which causes the production of a large number of different RDDs across jobs. 
Moreover, the probability of RDDs reaccess is low (e.g., the trace we generated has less than 26\% of RDDs are repeated across all jobs), and the temporal distances of RDDs reaccess are also relatively long~\cite{vcacheshare}. Thus, it becomes more challenging for a caching algorithm to make good caching decisions to reduce the total work under such a cache-unfriendly case.

\vspace{-3mm}
Fig.~\ref{FIG:EXP-SP-REC} shows the real experimental results under four different caching algorithms, i.e.,  FIFO, LRU, LCS, and Adaptive. To investigate the impact of cache size, we also change the size of storage memory pool to have different numbers of RDDs that can be cached in that pool.  
Compared to the other three algorithms, our adaptive algorithm  achieves non-negligible improvements on both hit ratio (see in Fig.~\ref{FIG:EXP-SP-REC}(a)) and makespan (see in Fig.~\ref{FIG:EXP-SP-REC}(b)), especially when the cache size increases. 
Specifically, the hit ratio can be improved by 13\% and the makespan can be reduced by 12\% at most, which are decent achievements for such a cache-unfriendly stress test with less room to improve performance. Furthermore, we observe that Adaptive significantly increases the hit ratio and reduces the makespan when we have more storage memory space, which again indicates that our caching algorithm has the ability to make good use of memory space. 
In contrast, the other algorithms have less improvement on hit ratio and makespan, since they cannot conduct cross-job computational overlap detection. While, with a global overview of all accessed RDDs, our adaptive algorithm  can effectively select proper RDDs from all jobs to be  cached in the limited storage memory pool.

\vspace{-3mm}
\section{Related Work}
\label{SEC:RW}
\vspace{-3mm}


\intechreport{In the era of big data, a large amount of data is needed to be analyzed and processed in a small amount of time. 
To meet the requirement, two types of in-memory processing systems are proposed~\cite{zhang2015memory}. %
The first type is data analytics system which is focusing on batch processing such as SINGA~\cite{singa}, Giraph~\cite{giraph}, and GridGain~\cite{gridgain}. 
The second type is real-time data processing systems such as Storm~\cite{storm}, Spark Streaming~\cite{sparkstream}, MapReduce Online~\cite{condie2010mapreduce}. 
}{}

Memory management is a well-studied topic across in-memory processing systems. 
Memcached~\cite{Memcache} and Redis~\cite{redis} are highly available distributed key-value stores.
Megastore~\cite{megastore} offers a distributed storage system with strong consistency guarantees and high availability for interactive online applications. 
EAD~\cite{yang2018ead} and MemTune~\cite{xu2016memtune} are dynamic memory managers based on workload memory demand and in-memory data cache needs. 
\intechreport{

A number of studies have also been done for modeling the multi-stage frameworks.  
Study~\cite{gu2013memory} compares the performance in both time and memory cost between Hadoop and Spark, and they observed that Spark is, in general, faster than Hadoop in iterative operations but Spark has to pay for more memory consumption. 
Study~\cite{wang2015performance} proposed a simulation driven prediction model that can predict job performance with high accuracy for Spark. 
A novel analytical model is designed in Study~\cite{wang2016modeling}, which can estimate the effect of interference among multiple Apache Spark jobs running concurrently on job execution time.

}{}
There are some heuristic approaches to evict intermediate data in big data platforms. 
%
Least Cost Strategy (LCS)~\cite{geng2017lcs} evicts the data which lead to minimum recovery cost in future. 
Least Reference Count (LRC) \cite{yu2017lrc} evicts the cached data blocks whose reference count is the smallest where the reference count dependent child blocks that have not been computed yet. 
Weight Replacement (WR)~\cite{duan2016selection} is another heuristic approach to consider computation cost, dependency, and sizes of RDDs.  
ASRW~\cite{wang2015new} uses RDD reference value to improve the memory cache resource utilization rate and improve the running efficiency of the program.
%
Study~\cite{kathpal2012analyzing} develops cost metrics to compare storage vs. compute costs and suggests when a transcoding on-the-fly solution can be cost-effective. 
Weighted-Rank Cache Replacement Policy (WRCP)~\cite{ponnusamy2013cache} uses parameters as access frequency, aging, and mean access gap ratio and such functions as size and cost of retrieval. 
These heuristic approaches do use optimization frameworks to solve the problem, but they are only focusing on one single job, and ignoring cross-job intermediate dataset reuse.

\vspace{-4mm}
\section{Conclusion}
\label{SEC:CO}
\vspace{-3mm}

The big data multi-stage parallel computing framework, such as Apache Spark, has been widely used to perform data processing at scale. 
To speed up the execution, Spark strives to absorb as much intermediate data as possible to the memory to avoid repeated computation. 
However, the default in-memory storage mechanism LRU does not choose reasonable RDDs to cache their partitions in memory, leading to arbitrarily sub-optimal caching decisions. 
In this paper, we formulated the problem by proposing an optimization framework, and then developed an adaptive cache algorithm to store the most valuable intermediate datasets in the memory.
According to our real implementation on Apache Spark, the proposed algorithm can improve the performance by reducing 12\% of the total work to recompute RDDs.
In the future, we plan to extend our methodology to support more big data platforms.
\vspace{-5mm}

\bibliographystyle{IEEEtran}
{\small \bibliography{07_References}}

\intechreport{\appendix
\subsection{Online Algorithm Overview}\label{app:overview}

 We describe here our adaptive algorithm for solving {\textsc{MaxCachingGain}} without a prior knowledge of the demands $\lambda_G$, $G\in \mathcal{G}$. The algorithm is based on \cite{ioannidis2016adaptive}, which solves a problem with a similar objective, but with matroid (rather than knapsack) constraints. We depart from \cite{ioannidis2016adaptive} in both the objective studied--namely, \eqref{obj}-- as well as in the rounding scheme used: the presence of knapsack constraints implies that a different methodology needs to be applied to round the fractional solution produced by the algorithm in each step.  
 
 We partition time into periods of equal length $T>0$, during which we collect access statistics for different RDDs. In addition, we maintain as state information the\emph{ marginals}  $y_v\in[0,1]^{|\mathcal{V}|}$: intuitively each $y_{v}$ captures the probability that node $v\in \mathcal{V}$ is cached. 
When the period ends, we (a) adapt the state vector $y=[y_v]_{v\in\mathcal{V}}\in [0,1]^{|\mathcal{V}|}$, and (b) reshuffle the contents of the cache, in a manner we describe below. 

\fussy
\noindent\textbf{State Adaptation.} We use RDD access and cost measurements collected  during a period to produce a random vector $z=[z_v]_{v\in \mathcal{V}}\in \reals_+^{|\mathcal{V}|}$ that is an unbiased estimator of a subgradient of $L$ w.r.t.~to $y$.  That is, if $y^{(k)}\in[0,1]^{|\mathcal{V}|}$ is the vector of marginals at the $k$-th measurement period, $z=z(y^{(k)})$ is a random variable satisfying:
 \begin{align}\label{estimateprop}
\expect\big[z(y^{(k)})\big] \in \partial L(y^{(k)})
\end{align}
where $\partial L(y)$ is the set of subgradients of $L$ w.r.t $y$.  
 We specify how to produce such estimates  below, in Appendix~\ref{app:distributedsub}. 
 
 Having these estimates, we adapt the state vector $y$ as follows: at the conclusion of the $k$-th period, the new state is computed as
 \begin{align}y^{(k+1)} \leftarrow \mathcal{P}_{\feasibledomain} \left( y^{(k)} + \gamma^{(k)}\cdot z(y^{(k)}) \right),\label{adapt}\end{align}
where $\gamma^{(k)}>0$ is a gain factor and $\mathcal{P}_{\feasibledomain}$ is the projection to the  set of relaxed constraints: 
$$\feasibledomain = \left\{y\in [0,1]^{|\mathcal{V}|} : \sum_{v\in \mathcal{V}}s_vy_{v}=K \right\}.$$
Note that $\mathcal{P}_{\mathcal{D}}$ is a projection to a convex polytope, and can thus be computed in polynomial time.

\noindent\textbf{State Smoothening.}
Upon performing the state adaptation \eqref{adapt}, each node $v\in V$ computes the following
``sliding average'' of its current and past states:
 \begin{align}\label{slide}\bar{y}^{(k)} = \textstyle \sum_{\ell = \lfloor\frac{k}{2} \rfloor}^{k} \gamma^{(\ell)} y^{(\ell)} /\left[\sum_{\ell=\lfloor \frac{k}{2}\rfloor}^{k}\gamma^{(\ell)}\right]   .\end{align}
 This ``state smoothening''  is necessary precisely because of the non-differentiability of $L$ \cite{nemirovski2005efficient}. Note that  $\bar{y}^{(k)} \in \feasibledomain$, as a convex combination of points in $\feasibledomain$.

\noindent\textbf{Cache Placement.} Finally, at the conclusion of a timeslot, the smoothened marginals $\bar{y}^{(k)}\in [0,1]^{|\mathcal{V}|}$ are \emph{rounded}, to produce a new integral placement $x^{(k)}\in\{0,1\}^{|\nodeset|}$ that satisfies the knapsack constraint \eqref{intcont}. There are several ways of producing such a rounding \cite{ageev2004pipage,swaprounding,kulik2009maximizing}. We follow the probabilistic rounding of \cite{kulik2009maximizing} (see also \cite{horel2014budget}): starting from a fractional $y$ that maximizes $L$ over $\feasibledomain$, the resulting (random) integral $x$ is guaranteed to be within $1-1/e$ from the optimal, in expectation. 

\subsection{Constructing an Unbiased Estimator of $\partial L(y)$.}\label{app:distributedsub}
To conclude our algorithm description, we outline here how to compute the unbiased estimates $z$ of the subgradients $\partial L(y{(k)})$ during a measurement period. In the exposition below,  drop the superscript $\cdot^{(k)}$ for brevity. 

The estimation proceeds as follows.
\begin{enumerate}
\item  Every time a job $G(V,E)$ is submitted for computation, we compute
the quantity
$$t_v=\textstyle\sum_{v\in V}\!c_v \id\big( x_v+\sum_{u\in \scc(v)}x_u\leq 1\big),$$
where 
$$
\id(A) =\begin{cases}
1, &\text{if}~A~\text{is true},\\
0, &\text{o.w.}
\end{cases}
$$
\item Let $\mathcal{T}_{v}$ be the set of quantities collected in this way at node $v$ regarding item $v\in \mathcal{V}$ during a measurement period of duration $T$. At the end of the measurement period, we produce the following estimates: \begin{align}z_{v}= \textstyle\sum_{t\in \mathcal{T}_{v} }t/T,\quad v\in\mathcal{V}.\label{estimation}\end{align} 
\end{enumerate}
Note that, in practice, $z_v$ needs to be computed only for RDDs $v\in \mathcal{V}$ that have been involved in the computation of some job in the duration of the measurement period. 

It is easy to show that the above estimate is an unbiased estimator of the subgradient: 
\begin{lemma}\label{subgradientlemma}
For  $z=[z_{v}]_{v\in \mathcal{V}}\in \reals_+^{|\mathcal{V}|}$ the vector constructed through coordinates \eqref{estimation},
$$\expect[z(y)] \in \partial L(y)\text{ and }\expect[\|z\|_2^2]<C^2|\mathcal{V}|^2(\Lambda^2+\frac{\Lambda}{T}),$$
where 
$C=\displaystyle\max_{v\in \mathcal{V}}c_v$ and  $\Lambda =\displaystyle\max_{v\in \nodeset}\!\!\sum_{G (V,E)\in \mathcal{G}: v\in V } \!\!\!\!\lambda_{G}.$
\end{lemma}
The proof of this lemma is almost identical, \emph{mutatis mutandis}, to the proof of Lemma 1 in \cite{ioannidis2016adaptive}. 

\subsection{Proof of Thm.~\ref{mainthm}.}\label{app:proofofmainthm}
To prove Thm.~\ref{mainthm}, we first have that, by 
setting $\gamma^{(k)}=\Theta(1/\sqrt{k})$,
\begin{align}\lim_{k\to\infty }\expect[L(\bar{y}^{k})] = \max_{y\in \mathcal{D}}L(y),\label{conv}\end{align}
and, in particular,
\begin{align}\lim_{k\to\infty}\prob[\bar{y}^{(k)}\in \argmax_{y\in \mathcal{D}} L(y) ] =1. \label{prob1}\end{align}
Eq.~\eqref{conv} is a consequence of Lemma~\ref{subgradientlemma} and Thm.~14.1.1, page 215 of Nemirofski~\cite{nemirovski2005efficient}. On the other hand, \eqref{sandwitch} implies that
\begin{align}F(\bar{y}^{(k)})\geq (1-1/e)  \max_{y\in \mathcal{D}}F(y)\label{lowerbound}\end{align}
for every $\bar{y}^{(k)}\in \argmax_{y\in \mathcal{D}} L(y),$ while the rounding scheme of \cite{kulik2009maximizing} ensures that the same property is attained by the rounded $x^{(k)}$ in expectation. This, along with \eqref{prob1}, implies the theorem.  
}{}
\end{document}